\documentclass[a4paper]{article}
\usepackage{color,graphicx}
\usepackage{amsmath}
\usepackage{amsthm}
\usepackage{amssymb}
\usepackage{amsfonts}
\RequirePackage{ifpdf} % running on pdfTeX?
\ifpdf
\usepackage[pdftex,bookmarks=true,
		   bookmarksnumbered=false,
		   bookmarksopen=false,
		   colorlinks=true,
		   linkcolor=webred] {hyperref}
\definecolor{webgreen}{rgb}{0, 0.5, 0} % less intense green
\definecolor{webblue}{rgb}{0, 0, 0.5} % less intense blue
\definecolor{webred}{rgb}{0.5, 0, 0}   % less intense red
\else
\newcommand{\href}[2]{ #1 }
\fi
\title{Netsukuku\\
{\small Close the world, \reflectbox{Open the next}}}
\author{http://netsukuku.freaknet.org}
\begin{document}
\maketitle

\begin{abstract}
Netsukuku is a P2P network system designed to handle a large number of nodes
with minimal CPU and memory resources. It can be easily used to build a
worldwide distributed, anonymous and not controlled network, separated from the
Internet, without the support of any servers, ISPs or authority controls.\\
In this document, we give a generic and non technical description of the
Netsukuku network, emphasizing its main ideas and features.
\end{abstract}

\section{The old wired}

The Internet is a hierarchic network managed by multinational companies and 
organisations supported by governments. Each bit of Internet traffic passes
through proprietary backbones and routers.\\
The Internet Service Providers give the connectivity to all the users, who
are in the lowest rank of this hierarchic pyramid. For this reason, the 
Internet is not a global network created by the users and shared between
them. The people can join the Net only in accordance with the restrictive
conditions and terms imposed by the Majors.

The Internet represents, today, the means to access information, knowledge
and communication. About 1 billion of people can connect to this great
proprietary network, but the remaining 5 billion of people, which don't have
enough economic resource, are still waiting the multinationals to supply a
service within their reach.

The Internet was born with the intent of warranting a secure and
unattackable communication between the various nodes of the network, but
now, paradoxically, an ISP has the power to cut out of the Internet entire
nations by simply stopping to give its services.

Beside that, Internet is not anonymous: the ISP can trace back and analyse
the traffic of data going through their servers, without any limits.

The centralised and hierarchical structure of the Internet creates, as a
consequence, other identical systems, based on it, i.e. the DNS.  
The servers of the Domain Name System are managed by different ISPs, as well
and the domains are literally sold through a similar centralised system.
This kind of structures allows, in a very simple and efficient way, to
physically localise any computers, which are connected to the Internet, in a
very short time and without any particular efforts.

In China, the whole net is constantly watched by several computers filtering
the Internet traffic: a Chinese will never be able to see or came to know
about a site containing some keywords, such as ``democracy'', censored by
the government. He'll never be able to express his own ideas on the net,
e.g. about his government's policy, without risking till the death
penalty.

Internet was born to satisfy the military needs of security for the
administration of the American defence, not to ensure freedom of
communication and information: in order to communicate between each other
the Internet users are obliged to submit themselves to the control and to
the support of big multinationals, whose only aim is to expand their own
hegemony.\\
As long as all the efforts to bring more freedom, privacy and accessibility
in the Internet face aversions, fears, contrary interests of governments and
private companies, the very alternative solution to this problem is to let
the users migrate toward a distributed, decentralised and fully efficient
net, in which all the users interact at the same level, with no privilege
and no conditioning means, as authentic citizens of a free world wide
community.

\section{The Netsukuku wired}

Netsukuku is a mesh network or a p2p net system that generates and sustains
itself autonomously. It is designed to handle an unlimited number of nodes
with minimal CPU and memory resources. Thanks to this feature it can be
easily used to build a worldwide distributed, anonymous and not controlled
network, separated from the Internet, without the support of any servers,
ISPs or control authorities.\\
This net is composed by computers linked physically each other, therefore
it isn't build upon any existing network. Netsukuku builds only the routes
which connects all the computers of the net.\\
In other words, Netsukuku replaces the level 3 of the model iso/osi with
another routing protocol.\\

Being Netsukuku a distributed and decentralised net, it is possible to
implement real distributed systems on it, e.g. the Abnormal Netsukuku Domain
Name Anarchy (ANDNA)\cite{andnadoc} which will replace the actual hierarchic
and centralised system of DNS.

\subsection{Gandhi}

The main characteristic of Netsukuku is its self-management: the network
dynamically configures itself without any external interventions.
All the nodes share the same privileges and limitations, giving their
contribution to sustain and expand Netsukuku.
The more the nodes increase in number the more the net grows and becomes
efficient.\\

The total decentralisation and distribution allows Netsukuku to be
neither controlled nor destroyed: the only way to manipulate or demolish
it, is to knock physically down each single node composing the network.

\subsection{No name, no identity}

Inside Netsukuku everyone, in any place, at any moment, can hook immediately
to the net without coming trough any bureaucratic or legal compliance.\\

Every elements of the net is extremely dynamic and it's never the same.
The IP address identifying a computer is chosen randomly, therefore it's
extremely difficult to associate it to a particular physical place; furthermore
since the routes are composed by a huge number of nodes, it becomes a titanic
task to trace a specific node.\\
The traffic of the nodes is protected by a complete cryptographic
layer\cite{carciofo}, which ensures a strong anonymity and security.

\subsection{So, what is it?}

Netsukuku is a mesh network built on top of its own dynamic routing protocol.

Currently there are a wide number of dynamic routing protocols, but they
are solely utilised to create small and medium nets.
The routers of Internet are also managed by different protocols such as the
OSPF, the RIP, or the BGP. They are based on different classical graph
algorithms, which are able to find out the best path to reach a node in a
given graph. However, all these protocols require a very high waste of CPU and
memory. Indeed, the Internet routers are computers specifically dedicated to
the executions of these algorithms. It would be impossible to implement one of
these protocols to create a maintain a mesh network large as the whole
Internet.

The Netsukuku protocol structures the entire net as a
fractal\cite{ntktopology} and, in order to calculate all the needed routes it
makes use of a particular algorithm called Quantum Shortest Path Netsukuku
(QSPN)\cite{qspndoc}.

A fractal is a structure which can be compressed infinitely, because every
part of itself is composed by the same fractal. Thus its high compression
level gives the ability to store the whole Netsukuku map in just few
Kilobytes.

On the other hand, the QSPN is an algorithm that has to be executed by the
network itself. The nodes, in order to execute it, have just to send and
receive the Tracer Packets, without using heavy computational resources.

\subsection{Netsukuku the wireless}
The best medium to establish physical connections between nodes is the wifi.
When Netsukuku will be widely adopted, its users will have just to set up their 
own wifi antennas on a well exposed place (i.e. windows or roofs), linking 
themselves to the other Netsukuku users, placed within their radio ranges.
Nowadays, there is a great variety of wifi technologies which allow to link
two nodes distant kilometres each other. A whole city can be easily covered by
placing just one node in each neighbourhood.\\
Moreover, it is possible to use virtual tunnels over the Internet to temporarily
replace missing physical links. This subject is described in \cite{inetdoc}.

%%%%%%%%%%%%%%%%
% Bibliography %
%%%%%%%%%%%%%%%%

\newpage

\begin{center}
\verb|^_^|
\end{center}
\end{document}